\documentclass[oneside,reviewcopy,10pt]{elsart}   %main format that used
\usepackage{graphicx}  %package needed
\usepackage{subfigure}

\usepackage{amssymb}
\usepackage{threeparttable}

\begin{document}
\bibliographystyle{unsrt}
\begin{frontmatter}             %front page

\title{A Hermite pseudospectral solver for two-dimensional incompressible flows on infinite domains}

\author{Zhaohua Yin}
\address{National Microgravity Laboratory, Institute of Mechanics, Chinese Academy of Sciences, No.15 Beisihuanxilu, Beijing 100190, P.R. China}
\ead{zhaohua.yin@gmail.com; zhaohua.yin@imech.ac.cn}

\begin{abstract}   %abstract, rewrite
The Hermite pseudospectral method is applied to solve the Navier-Stokes equations on a two-dimensional infinite domain. The feature of Hermite function allows us to adopt larger time steps than other spectral methods, but also leads to some extra computation when the stream-function is calculated from the vorticity field. The scaling factor is employed to increase the resolution within the region of our main interest, and the aliasing error is fully removed by the 2/3-$rule$. Several traditional numerical experiments are performed with high accuracy, and some related future work on physical applications of this program is also discussed.
\end{abstract}

\end{frontmatter}

\section{Introduction}

In practice, the boundary effect of flows is inevitable and requires careful treatment. On the other hand, when people concentrate on physical mechanism of fluid, they hope that the boundary effect can be fully removed. For example, the Oseen vortex in two-dimensional(2D) fluid adopts the assumption of the infinite domain (e.g., see \cite{Montgomery2011} and references therein). It has also been shown in many previous investigations on how the boundary can influence the physical mechanism of flows. Even when the region of main interest are fairly far away from the solid boundary, the drop speeds show about 15\% difference in a thermocapillary migration study\cite{Yin2012}. In the research of the axisymmetrization of an isolated 2D vortex\cite{McWilliams1987}, a double-sized computing domain is adopted to alleviate the vortex overrotation caused by the periodical boundary in a Fourier spectral simulation\cite{Platte2009}. Hence, it is essential to adopt the infinite domain in such studies. When the infinite computing domain ($-\infty$, $+\infty$) is considered, one possible way is still using the finite-domain solver but with a certain mapping between ($x_{min}$, $x_{max}$) and ($-\infty$, $+\infty$). Another way is something like the sponge-layer suggested in \cite{Mariotti1994}, which absorbs the incoming vortex filaments. The other way is more natural: Laguerre or Hermite functions are adopted to construct global approximation to functions defined on unbounded intervals.

The normalized Hermite function of degree $n$ is defined as:
\begin{equation}
\label{hermite}
\hat {H}_n (x) = \frac{1}{\sqrt {2^nn!} }e^{ - \frac{x^2}{2}}H_n (x),
\end{equation}
where $H_n (x)$ are the usual (unnormalized) Hermite polynomials (for instance, see \cite{Shen2006,Shen2009}). ${\hat {H}_n (x)}$ are orthogonal in $L^2$($-\infty$, $\infty$):
\[
\int^{+\infty}_{-\infty} \hat{H}_k(x) \hat{H}_m(x) = \sqrt{\pi} \delta_{km}, \; k,m \geq 0.
\]

In the last few decades, many investigations concerning the theory and application of Hermite functions have been carried out, and an early review can be seen in \cite{Boyd2001}. Although the computing domain is infinite for Hermite spectral methods, the region of our main interest is still finite. This fact led to the appearance of the most important concept in the practical sense: the scaling factor\cite{Tang1993,Funago1991}.  There have been many Hermite spectral (or, pseudospectral) studies which investigated partial differential equations from different fields \cite{Tang1992,Guo1999,Guo2000,Fok2001,Schumer1998,Weishaupl2007,Parand2010,Parand2011,Hagedorn1998,Kieri2012}. Most researches in this field have so far dealt with one-dimensional problems, or multi-dimensional problems with only one Hermite direction. One most recent paper \cite{Xiang2013} deals with two-dimensional partial differential equations with two Hermite directions, and takes an elliptic equation with a harmonic potential and a class of nonlinear wave equations into consideration. However, so far as we know, there has been no effort in solving the multi-dimensional Navier-Stokes (NS) equations with pure Hermite spectral (or Hermite pseudospectral) methods, and this is the main object of this paper.

The paper is arranged as follows: the numerical details are described in Sec. 2, and the performance of some validating tests is introduced in Sec. 3, and the discussions and some related future work are presented in Sec. 4.

\section{Numerical scheme for the Hermite pseudospectral NS solver}

The 2D incompressible NS equations on the infinite domain $\vec{x}=(x,y)\in [-\infty, +\infty] \times [-\infty, +\infty]$ in terms of vorticity and stream function are written as
\begin{equation}
\label{eqome} \frac{\partial \omega }{\partial t} + {\rm {\bf u}}
\cdot \nabla \omega = \nu \Delta \omega ,
\end{equation}
\begin{equation}
\label{eqpsi} \Delta \psi = - \omega ,
\end{equation}
where ${\rm {\bf u}} = (u, \upsilon)$ is the velocity, $\nu$ the kinematic viscosity, $\psi$ stream function, and vorticity ${\rm {  \mbox{ \boldmath{$\omega$}}}} = (0,0,\omega ) = \nabla \times {\rm {\bf u}}$. The stream function is related to velocity by $u = {\partial \psi }/{\partial y}$ and  $\upsilon = -{\partial \psi }/{\partial x}$. We use pseudospectral methods to solve Eq.~(\ref{eqome})  by expanding $u$, $\upsilon$, $\psi$, and $\omega$ in a truncated Hermite series.

\subsection{Expansions}

\begin{figure}
\centering
\scalebox{1}[1]{\includegraphics[width=\linewidth]{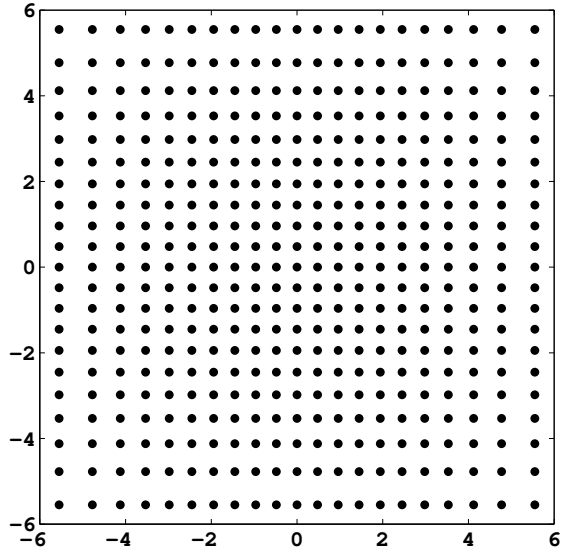}}
\caption{\label{fig:grids} Distribution of grid points for the resolution of $20 \times 20$. Here, $x_0 = 5.55$ is the largest root of $\hat {H}_{21} (x)$. Note that the density of points is slightly higher in the center, which is much clearer in Fig. \ref{fig:ngrids}.}
\end{figure}

In pseudo-spectral methods, the optimal pseudo-scpectral points are the roots
of $\hat {H}_{N + 1} (x)$, which are denoted by $\{x\}_{j=0}^N$ with the order $x_0 > x_1 > \ldots > x_N$. Since $x_0 = -x_N \approx \sqrt{2N}$, the values of vorticity in our solver are defined on the grid points in the box of [-$x_0$,$x_0$]$\times$[-$x_0$,$x_0$] (e.g., see Fig. \ref{fig:grids} or Figs. 6 in~\cite{Shen2009}).

\begin{figure}
\centering
\scalebox{0.9}[0.9]{\includegraphics[width=\linewidth]{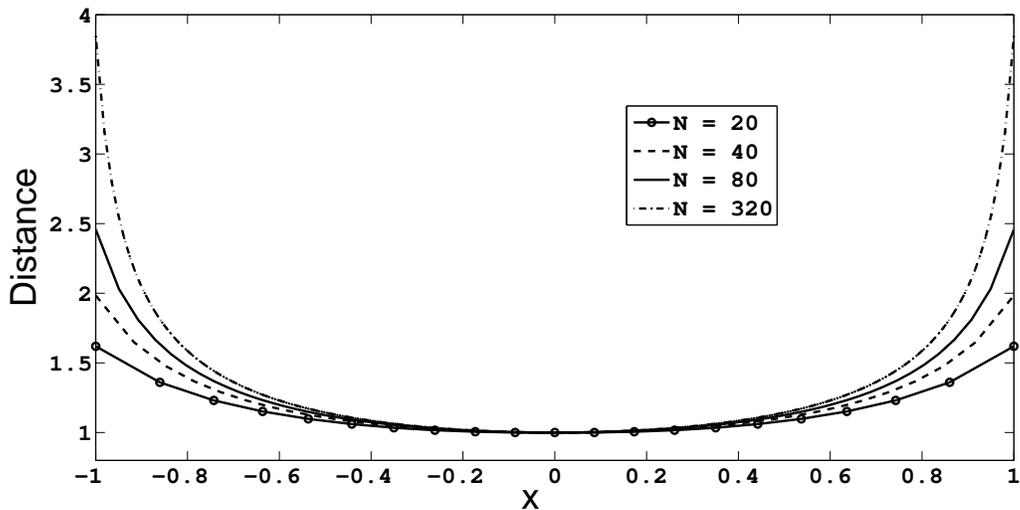}}
\caption{\label{fig:ngrids}  Distances between grid points and their neighboring points for different resolutions. All points are confined within the $[-1,1]$ box. The distance is scaled by the shortest distance of the corresponding resolution, which always locates in the center.}
\end{figure}

With the three-term recurrence:
\begin{eqnarray}
&\hat{H}_0(x) = e^{-x^2/2}, \;  \hat{H}_1(x) = \sqrt{2} x e^{-x^2/2}, \nonumber \\
&\hat{H}_{n+1}(x) = x \sqrt{\frac{2}{n+1}}\hat{H}_n(x) - \sqrt{\frac{n}{n+1}}\hat{H}_{n-1}(x), \; n\geq1,
\end{eqnarray}
the values of $\hat{H}_n(x_j)$, $0 \leq n,j \leq N$ can be calculated. The 2D vorticity field is then transformed to the Hermite spectral space with the following equation:
\begin{equation}
\label{eqback}
\omega (x,y) = \sum\limits_{k_x = 0}^N \sum\limits_{k_y = 0}^N \hat {\omega }_{k_x,k_y} \hat {H}_{k_x} (x)\hat {H}_{k_y} (y),
\end{equation}
where $\hat {\omega }_{k_x, k_y} $is the Hermite coefficients and $\textbf{k} = (k_x,k_y)$ wave numbers for $x$ and $y$ directions.
The Hermite coefficients are determined by
\begin{equation}
\label{eqfor}
\hat{\omega}_{k_x,k_y} = \frac{\pi}{(N+1)^2} \sum^N_{j=0}
\frac{\hat H_{k_y}(y_j)}{\hat H_N^2(y_j) }
\left(
\sum^N_{i=0} \frac{\hat H_{k_x}(x_i)}{\hat H_N^2(x_i) } \omega(x_i,y_j)
\right),
\; \; \;
0 \leq k_x,k_y \leq N.
\end{equation}
The above direct evaluation takes $O(N^3)$ operations. With the Fast Hermite Transform~\cite{Driscoll1997,Leibon2008}, only $O(N^2log^2N)$ operations are needed.

In real simulations, it is impossible to focus on the infinite domain, so we always concentrate on some small regions, e.g. $[-L,L]\times[-L,L]$. Hence, the scaling factor $a=\gamma_0/L$ is used in Eq. (\ref{eqback}) since
\[
\omega (x,y) = \sum\limits_{k_x = 0}^N \sum\limits_{k_y = 0}^N \hat {\omega
}_{k_x,k_y} \hat {H}_{k_x} (ax)\hat {H}_{k_y} (ay)
\]
is equivalent to
\[
\omega (x/a,y/a) = \sum\limits_{k_x = 0}^N \sum\limits_{k_y = 0}^N \hat {\omega
}_{k_x,k_y} \hat {H}_{k_x} (x)\hat {H}_{k_y} (y).
\]
In this paper, the determination of $L$ is 1) as small as possible; 2) to make sure that the vorticity outside the box of grid points is zero.

\subsection{Differentials}

The Hermite coefficients of first derivatives can be obtained through those of the primitive function with $O(N^2)$ operations, for example:

 for all $0\leq k_x,\; k_y \leq N$,
\[
\hat{(\omega_x)}_{k_x,k_y} =  \sqrt{\frac{k_x+1}{2}}\hat{\omega}_{k_x+1, k_y} - \sqrt{\frac{k_x}{2}} \hat{\omega}_{k_x-1, k_y},
\]
\[
\hat{(\omega_y)}_{k_x,k_y} =  \sqrt{\frac{k_y+1}{2}}\hat{\omega}_{k_x, k_y+1} - \sqrt{\frac{k_y}{2}} \hat{\omega}_{k_x, k_y-1},
\]
with the understanding that $\hat{\omega}_{-1, j} = \hat{\omega}_{j, -1}= \hat{\omega}_{N+1, j} = \hat{\omega}_{j, N+1} =0$, for $0 \leq j \leq N$. And the coefficients of the higher-order derivatives can be obtained similarly from those of lower-order derivatives.

\subsection{Solving the Navier-Stokes Equations: Eq.~(\ref{eqpsi})}

The reverse of $\psi$ from $\omega$ by Eq. (\ref{eqpsi}) is quite expensive since
\begin{eqnarray}
\label{poisson}
& - \hat{(\omega)}_{k_x,k_y} =  \hat{(\psi_{xx})}_{k_x,k_y} + \hat{(\psi_{yy})}_{k_x,k_y} = & \; \\
& \frac{\sqrt{(k_x+1)(k_x+2)}}{2}\hat{\psi}_{k_x+2, k_y}
- \frac{2k_x+1}{2} \hat{\psi}_{k_x, k_y}
+ \frac{\sqrt{k_x(k_x-1)}}{2}\hat{\psi}_{k_x-2, k_y}  & \nonumber \\
& +  \frac{\sqrt{(k_y+1)(k_y+2)}}{2}\hat{\psi}_{k_x, k_y+2}
- \frac{2k_y+1}{2} \hat{\psi}_{k_x, k_y}
+ \frac{\sqrt{k_y(k_y-1)}}{2}\hat{\psi}_{k_x, k_y-2}. &
\end{eqnarray}

A sparse $(N+1)^2 \times (N+1)^2$ matrix with $5N(N+8)$ non-zero elements has to be dealt with. Our numerical experiments show that the resultant equation system is not very well conditioned, and some pivoting is necessary for fairly large $N$ to solve the above equations (e.g., see~\cite{Press2002}).

\subsection{Solving the Navier-Stokes Equations: Eq.~(\ref{eqome})}
\label{solving}
The actual time integration of Eq.~(\ref{eqome}) is carried out by evaluating the Hermite expansion coefficients in course of time, and we only go back to physical space when we compute the nonlinear term $J={\bf {u}}\cdot{\bf {\nabla}}{\omega}$ of the vorticity equation. We discretize the advection term $J$ with a $2^{nd}$ order Adams-Bashforth scheme. The application of Eq.~(\ref{eqome}) in spectral space gives
\begin{equation}
{\hat{\omega}}_{\bf{k}}^{t+1} - {\hat{\omega}}_{\bf{k}}^{t} =
-\Delta t (\frac{3}{2}{\hat{J}}_{\bf{k}}^{t} - \frac{1}{2}{\hat{J}}_{\bf{k}}^{t-1}) -
\nu \Delta t{\bf{k}}^2{\hat{\omega}}_{\bf{k}}^{t},
\label{discret}
\end{equation}
where ${\hat{\omega}}_{\bf{k}}^{t}$ and ${\hat{J}}_{\bf{k}}^{t}$ are the Hermite coefficients of $\omega$ and $J$ at time $t$, respectively. The value of ${\hat{J}}_{\bf{k}}^{t}$ is obtained by the combination of Hermite transforms and the so-called de-aliasing technique by padding or truncation~\cite{can87}.

\vspace{0.2in}

As a comparison, the time integration for Fourier or Chebyshev spectral solver normally adopts a semi-implicit scheme. For example, the so-called ABCN scheme is adopted in a Fourier code~\cite{Yin2004}:

\begin{equation}
{\hat{\omega}}_{\bf{k}}^{t+1} - {\hat{\omega}}_{\bf{k}}^{t} =
-\Delta t (\frac{3}{2}{\hat{J}}_{\bf{k}}^{t} - \frac{1}{2}{\hat{J}}_{\bf{k}}^{t-1}) -
\frac{\nu \Delta t}{2}{\bf{k}}^2({\hat{\omega}}_{\bf{k}}^{t+1} +
{\hat{\omega}}_{\bf{k}}^{t})~.
\label{discret2}
\end{equation}
There are two reasons why the explicit time scheme is adopted in this solver:
\begin{itemize}
\item
It has been shown that the spectral radii for the first and second Hermite differentiation matrices are $O(\sqrt{N})$ and O$(N)$, respectively~\cite{Weideman1992}. This places rather weak stability restrictions on the Hermite method, and for the standard heat equation~\cite{Shen2006}, a maximum step size in the time direction of order $O(1/N)$ is required, whereas for Fourier and Chebyshev methods it is of order $O(1/N^2)$, $O(1/N^4)$, respectively. In the actual calculations this means that we need not even consider implicit time
integration methods with the Hermite method.

\item
For the Fourier spectral methods, the solution of Poisson equation can be obtained with trivial efforts of $O(N^2)$ operations, so the semi-implicit scheme only causes few extra computations. On the other hand, the semi-implicit scheme of the Hermite solver leads to an equation system similar to that of Eq.~(\ref{poisson}), which means extra nontrivial computation and computer memory.  Our numerical experiment shows that the adoption of the semi-implicit scheme in Hermite simulations does not lead to a larger time step.
\end{itemize}

\section{Validation and comparison}

The main purpose of this section is to validate the Hermite pseudospectral solver, and to show some improvement of it when compared with other numerical schemes. Three numerical experiments are performed:
\begin{itemize}
\item
a steady axisymmetric solution for 2D NS equation (subsection 3.1);
\item
an unsteady axisymmetric solution with exact formula to describe its evolution (subsection 3.2);
\item
an unsteady non-axisymmetric solution without any exact formula to describe its evolutions, but its comparison is also easy since it has been widely adopted for validating new codes in many previous studies (subsection 3.3).
\end{itemize}

\begin{table}
\centering
\caption{The error in the Burgers vortex simulations at $t=4.0$. }
\label{tab1}
\begin{tabular}{c|c|c}
  \hline   \hline
  Time step & Resolution & $L_\infty$ error \\
  \hline
  0.0005 & 120x120 & $1.03 \times 10^{-3}$ \\
  \hline
  0.00025 & 200x200 & $5.62 \times 10^{-4}$ \\
  \hline
  0.000125 & 400x400 & $2.70 \times 10^{-4}$ \\
  \hline   \hline
\end{tabular}
\end{table}

\subsection{Burgers vortex: a steady solution}

The Burgers vortex is a steady viscous vortex maintained by a secondary flow, and provides an excellent example of a balance between convection, intensification and diffusion of vorticity. In the 2D case, the corresponding governing equations are
\begin{equation}
\label{eqburgers}
\frac{\partial \omega }{\partial t} + (-\alpha x + u)\frac{\partial \omega }{\partial x}
+ (-\beta y + \upsilon)\frac{\partial \omega }{\partial y} = (\alpha + \beta)\omega + \nu \Delta \omega,
\end{equation}
\begin{equation}
\label{eqburgers1}
\Delta \psi = -\omega, \; u=\frac{\partial \psi}{\partial y} , \; \upsilon=-\frac{\partial \psi}{\partial x}.
\end{equation}
When $\alpha = \beta > 0$, $\omega(x,y) = \Omega e^{-{\alpha (x^2+y^2)}/{2\nu}}$ is an exact steady solution to the above equations~\cite{Robinson1984}, where the constance $\Omega$ is arbitrary and also a measure of the magnitude of the vorticity.

In the following simulation, the grid points are confined in a box of $[-3.5,3.5] \times [-3.5,3.5]$. For other parameters, we choose $\Omega = 10.0$, $\alpha = 0.012$ and $\nu = 0.0025$.

To justify the result of our simulation, the $L_\infty$ error is introduced as
\[L_\infty = \mbox{max} \left( \mbox{abs} \left( \frac{\omega_{simulation}(x,y,t)-\omega_{exact}(x,y,t)}{\Omega} \right) \right), \]
and the above definition will also be adopted in the next subsection.

Three different resolutions are adopted in this subsection. The errors between the $t=4.0$ results from our simulation and the original initial input data are shown in Tab.~\ref{tab1}. It seems that, for all simulations, the $L_\infty$ errors are very small, and decrease with higher resolutions.

As discussed before in subsection~\ref{solving}, the time steps for different grids in Tab.~\ref{tab1} show that the maximum value of time step is of order $O(1/N)$, even with the explicit time integration methods as described in Eq. (\ref{discret}). The same conclusion can be drawn for all simulations described in this section. The only difference in this subsection is that all time steps here are much smaller that those in Tab.~\ref{tab2} because of the existence of the secondary flow (see the $\alpha$ and $\beta$ related terms in Eq. (\ref{eqburgers})).

\subsection{Lamb-Oseen vortex: a self-similar unsteady solution}

Eqs. (\ref{eqburgers})\&(\ref{eqburgers1}) are not real since they are maintained by a secondary flow. Without any external force or second flow, there will be no steady vortex for Eqs. (\ref{eqome})\&(\ref{eqpsi}). However, there is an exact time-dependent solution for the initial condition $\omega(x,y,0) = 2\pi \exp(-(x^2+y^2))$:
\begin{equation}
\label{exact}
\omega(x,y,t) = \frac{2\pi}{1+4 \nu t}  \exp(-\frac{x^2+y^2}{1+4 \nu t}).
\end{equation}
This vortex is named after Horace Lamb and Carl Wilhelm Oseen (the Lamb-Oseen vortex)~\cite{Saffmann1992}, which models a line vortex that decays due to viscosity. Note that this solution is still axisymmetric and self-similar, but can provide more confirmation about our explicit time integration methods when compared with the steady solution in the previous subsection.

In the following simulations, the grid points are confined in a box of $[-2 \pi, 2 \pi] \times [-2 \pi, 2 \pi]$, and viscosity is set to be 0.00037. Three different resolutions are adopted. The errors between the $t=4.0$ results from our simulations and Eq. (\ref{exact}) are shown in Tab.~\ref{tab2}. Again, for all simulations, the $L_\infty$ errors are very small, and decrease with higher resolutions.

\begin{table}
\centering
\caption{The error in the Lamb-Oseen vortex simulations at $t=4.0$.}
\label{tab2}
\begin{tabular}{c|c|c}
  \hline   \hline
  Time step & Resolution & $L_\infty$ error \\
  \hline
  0.0025 & 120x120 & $1.38 \times 10^{-3}$ \\
  \hline
  0.00125 & 200x200 & $6.54 \times 10^{-4}$ \\
  \hline
  0.000625 & 400x400 & $2.32 \times 10^{-4}$ \\
  \hline   \hline
\end{tabular}
\end{table}

\begin{figure}
\centering
\scalebox{1}[1]{\includegraphics[width=\linewidth]{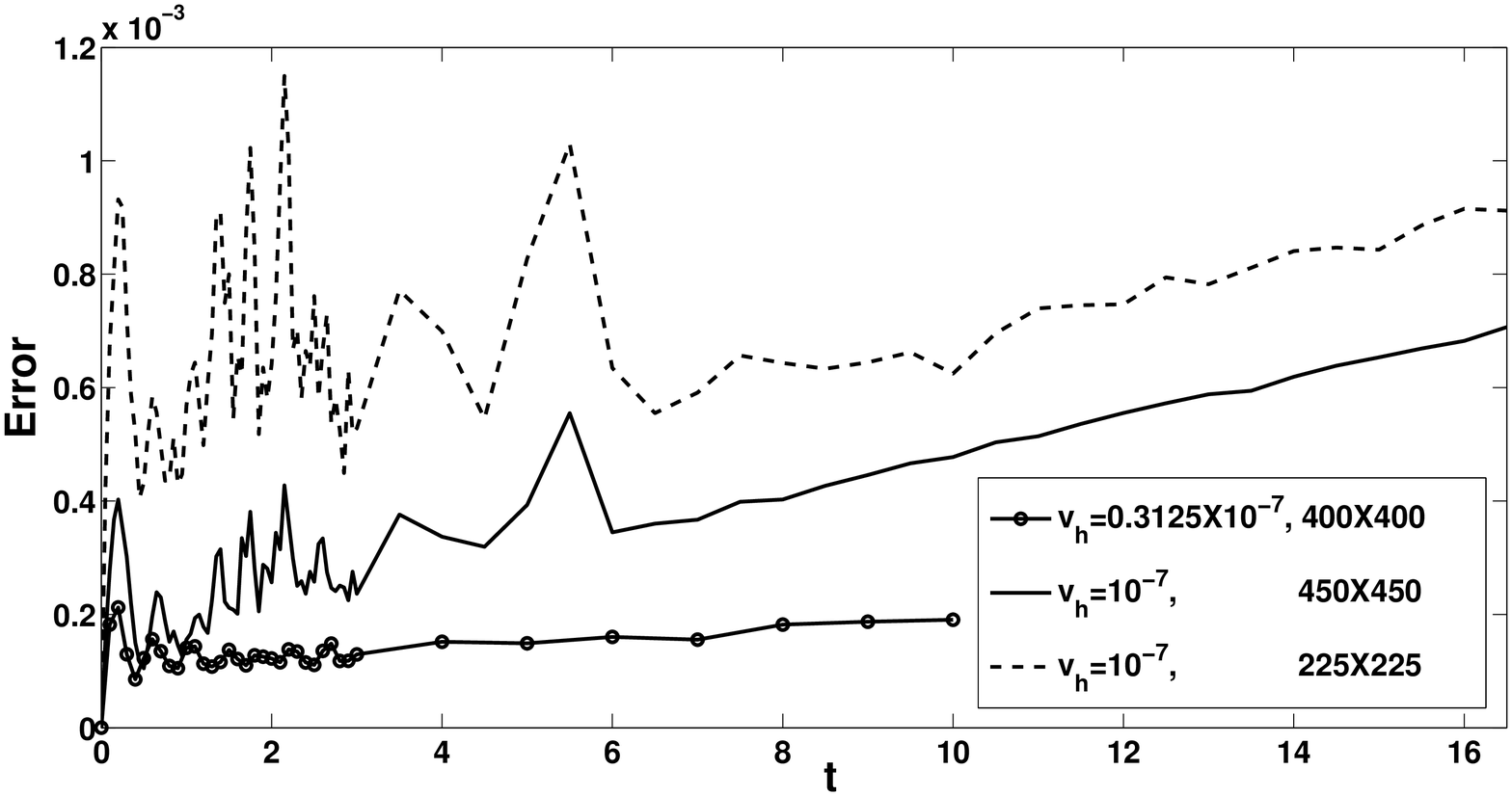}}
\caption{\label{fig:Merror} The relative errors in the conservation of maximum vorticity for the three Hermite simulations in subsection 3.3.}
\end{figure}

\begin{figure}
\center
\scalebox{0.6}[0.6]{\includegraphics[width=\linewidth]{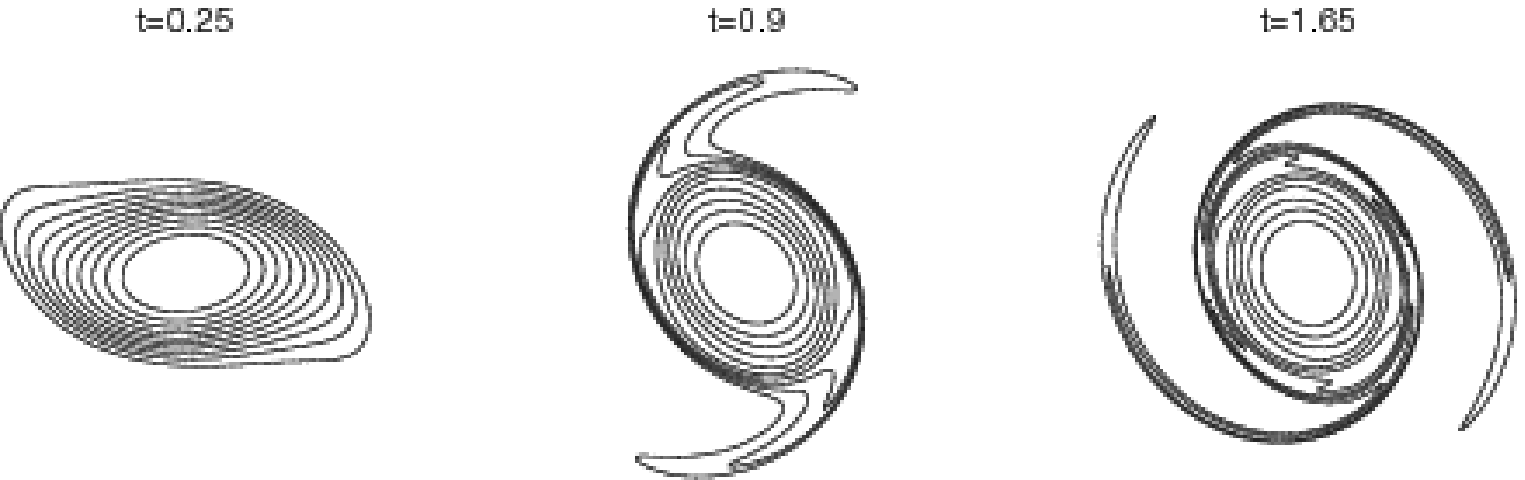}}
\scalebox{0.6}[0.6]{\includegraphics[width=\linewidth]{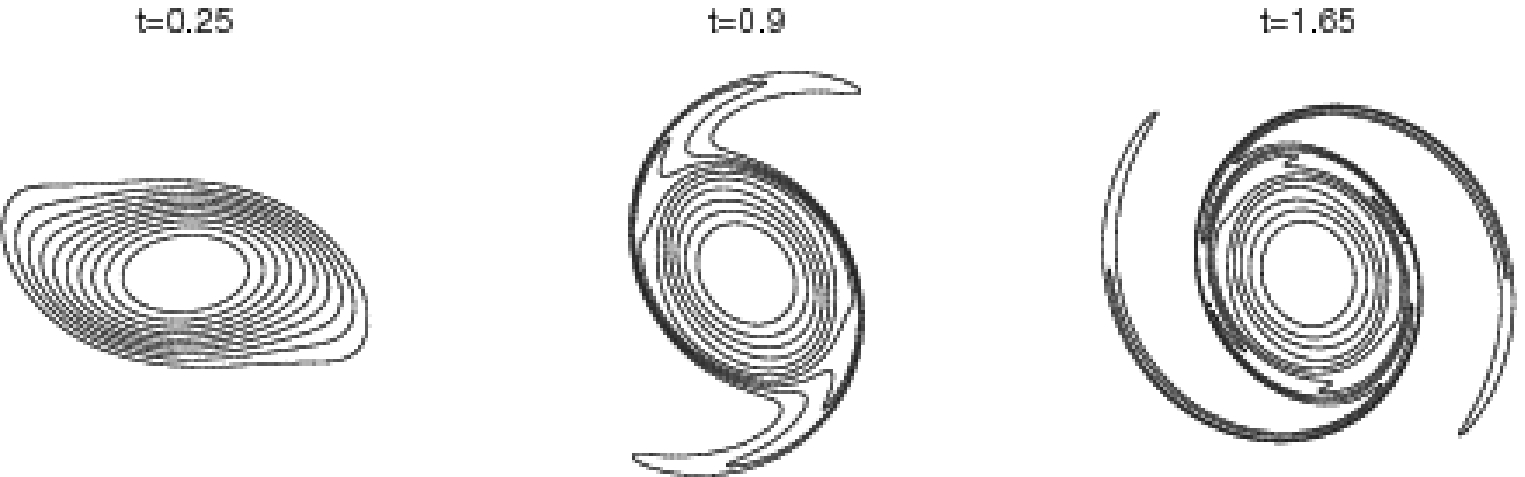}}
\caption{\label{fig:sdsize} Contours of vorticity for Run1 (the first row) and Run2 (the second row). All plots are rotated by 124.6 degree to compare with the results of Figs. 5.2 in \cite{Platte2009} with the vortex methods.}
\end{figure}

\begin{figure}
\centering
\scalebox{0.8}[0.8]{\includegraphics[width=\linewidth]{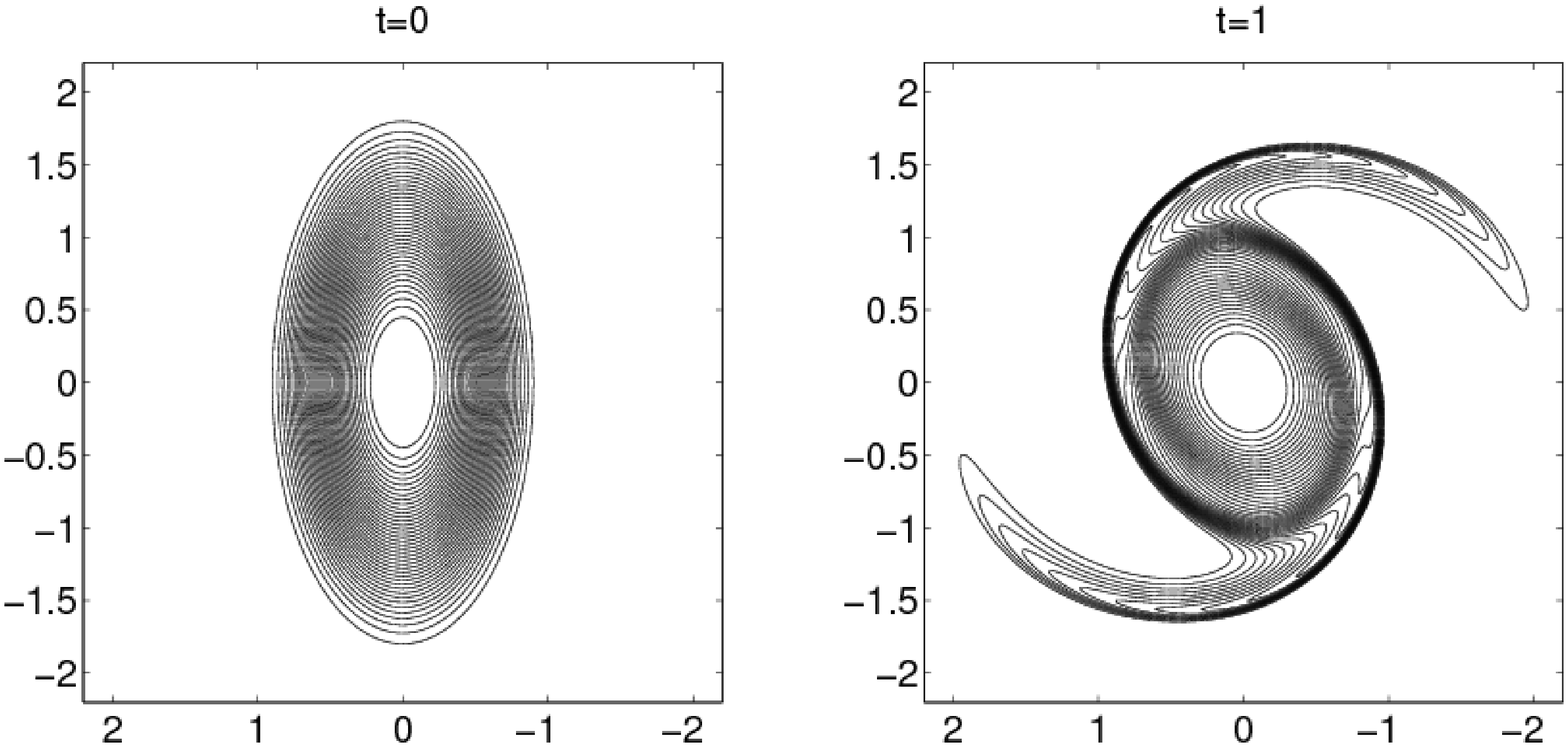}}
\scalebox{0.8}[0.8]{\includegraphics[width=\linewidth]{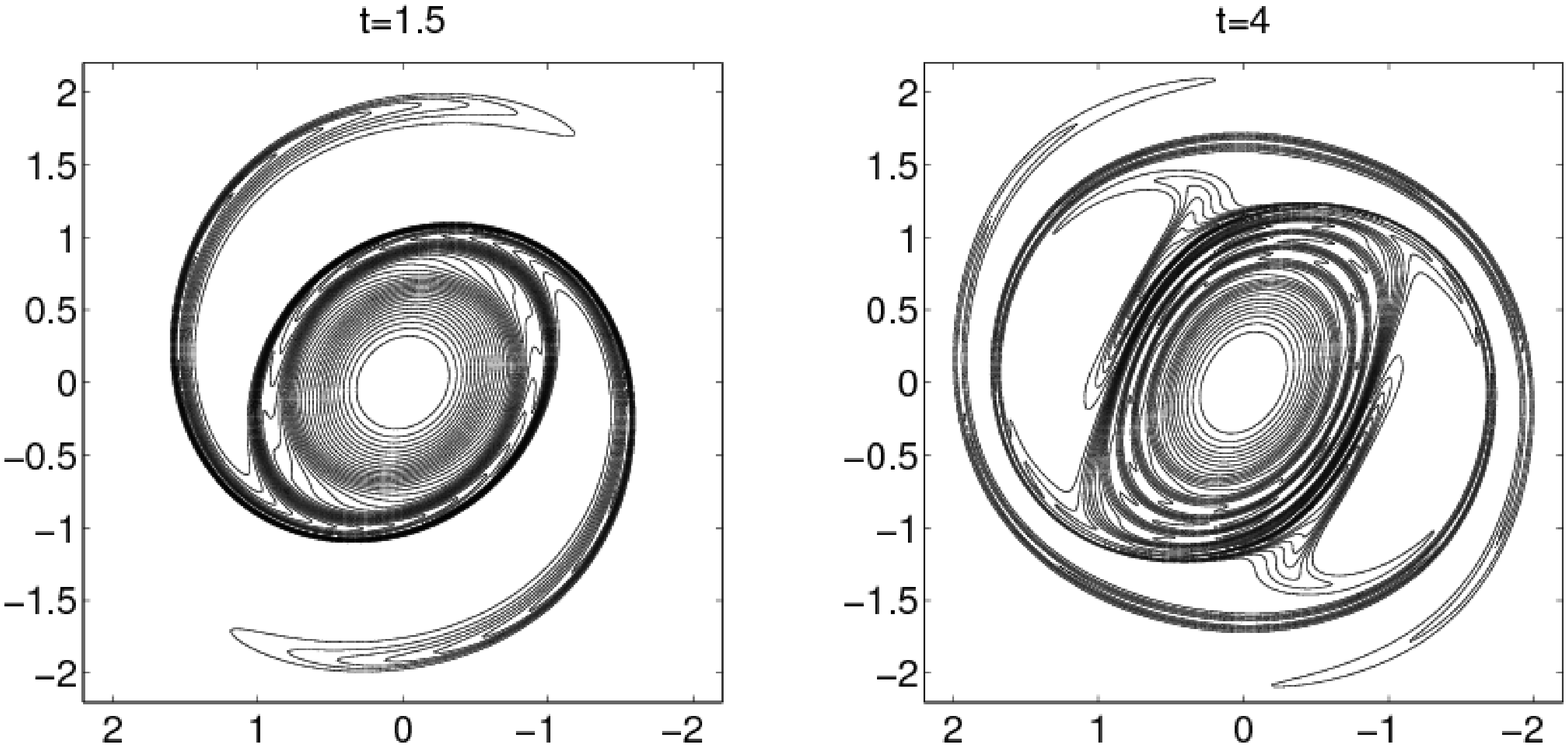}}
\scalebox{1.0}[1.0]{\includegraphics[width=\linewidth]{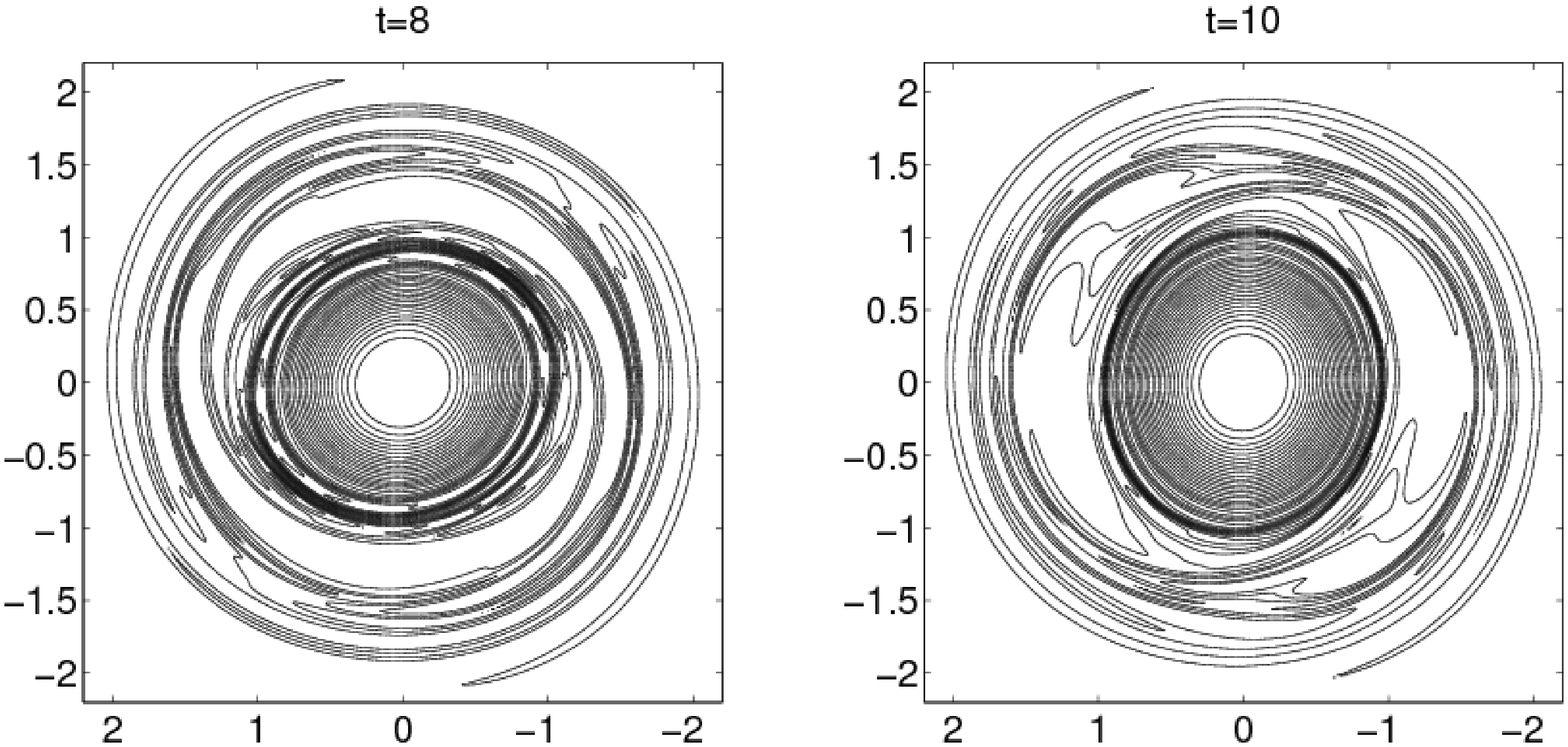}}
\caption{\label{fig:sumupn} Contours of vorticity for Run3, and all plots are rotated by 90 degree to compare with the vortex method results of Figs. 10 in \cite{Koumoutsakos1997}.}
\end{figure}

\begin{figure}
\centering
\begin{minipage}[c]{.7 \linewidth}
\centering
\scalebox{0.4}[0.4]{\includegraphics[width=\linewidth]{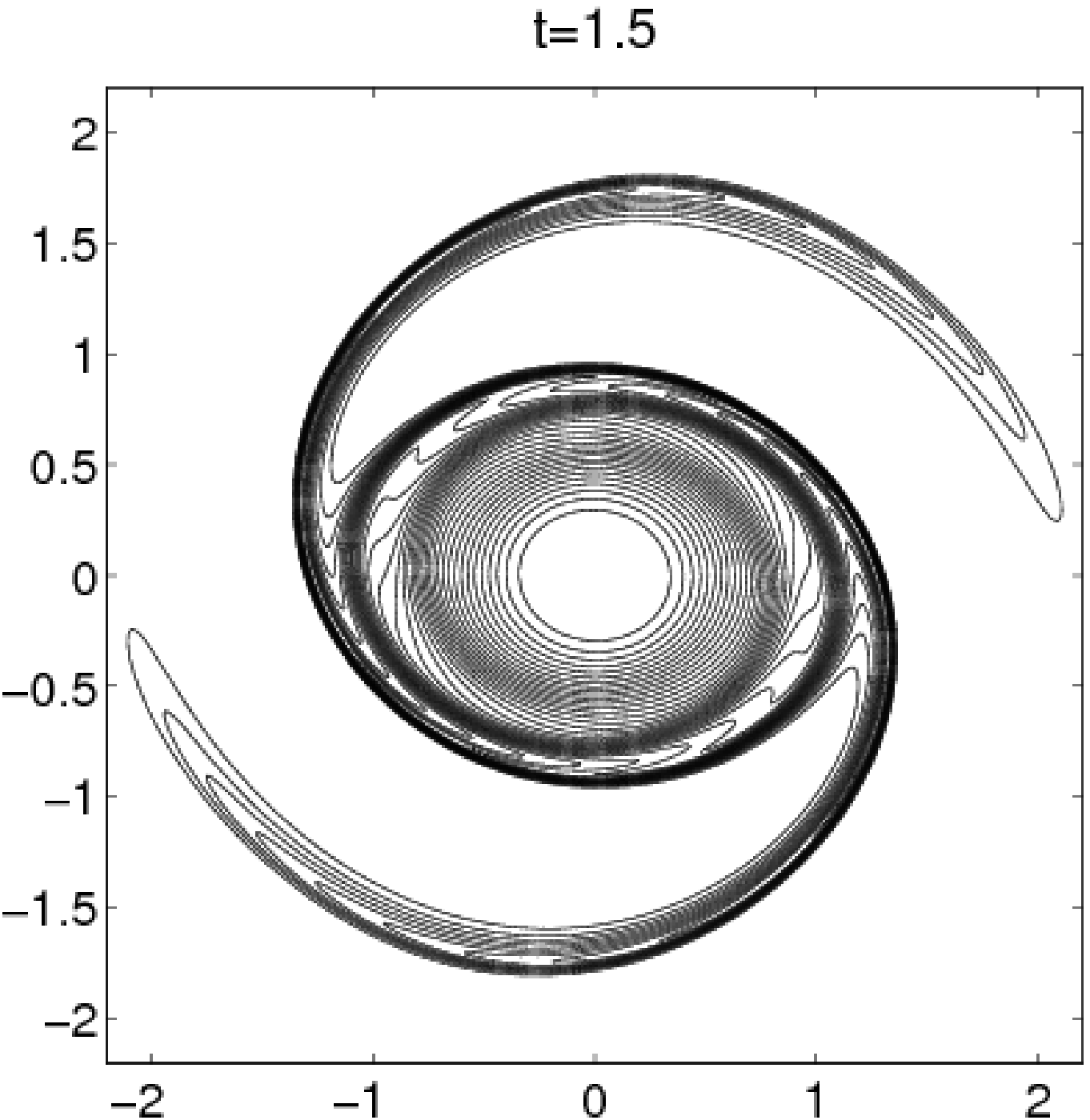}}
\scalebox{0.4}[0.4]{\includegraphics[width=\linewidth]{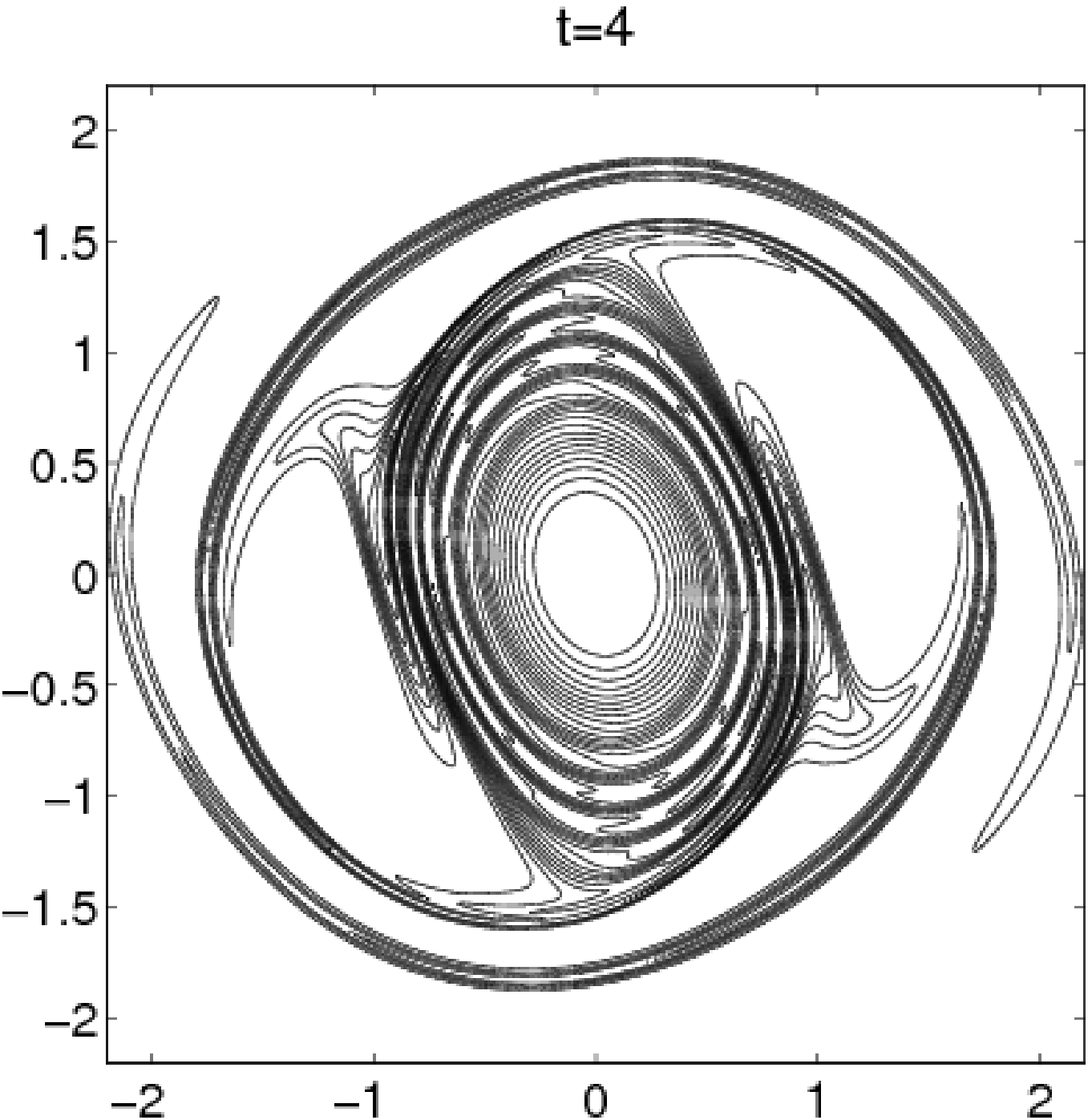}}
\scalebox{0.4}[0.4]{\includegraphics[width=\linewidth]{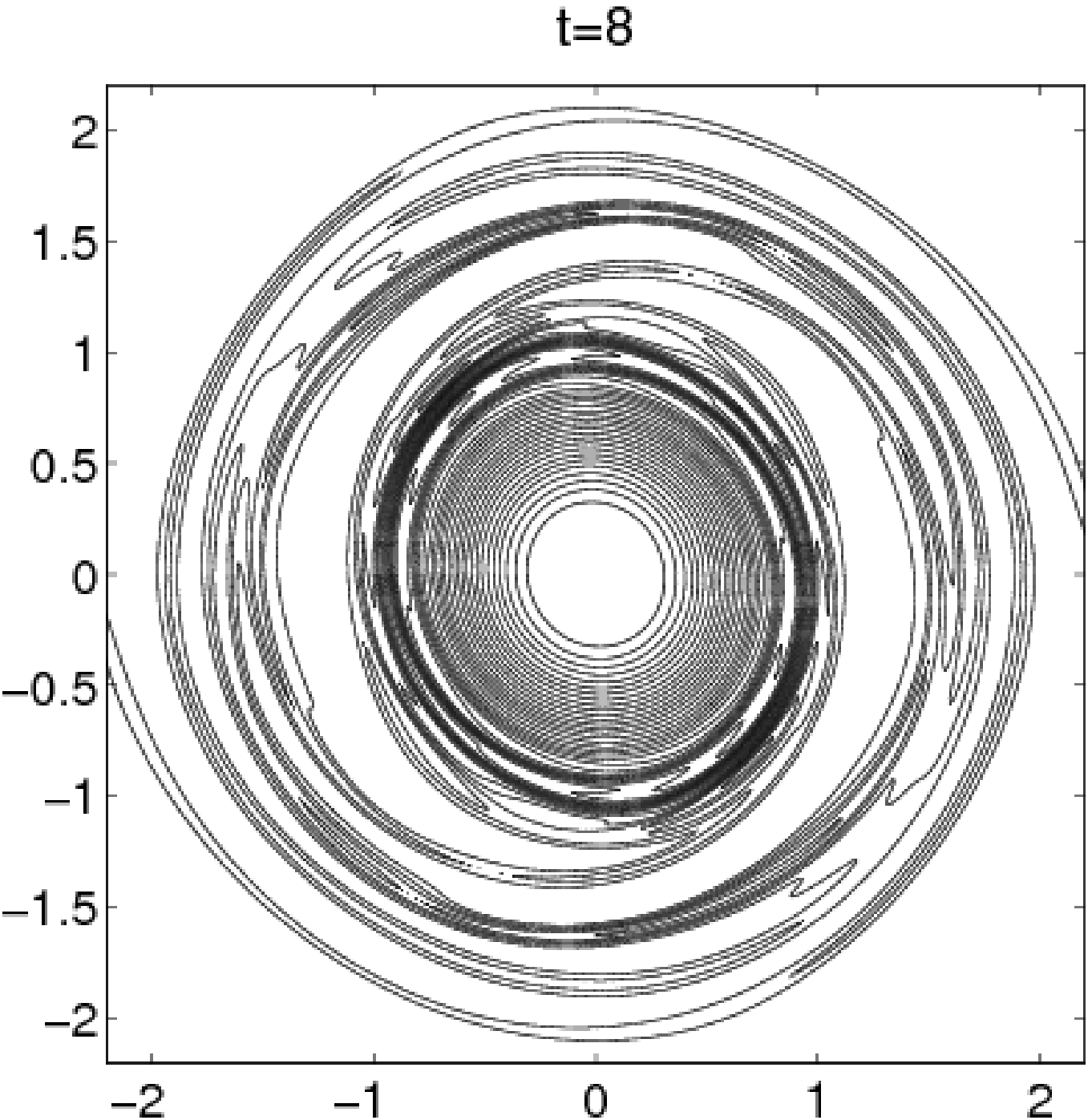}}
\scalebox{0.4}[0.4]{\includegraphics[width=\linewidth]{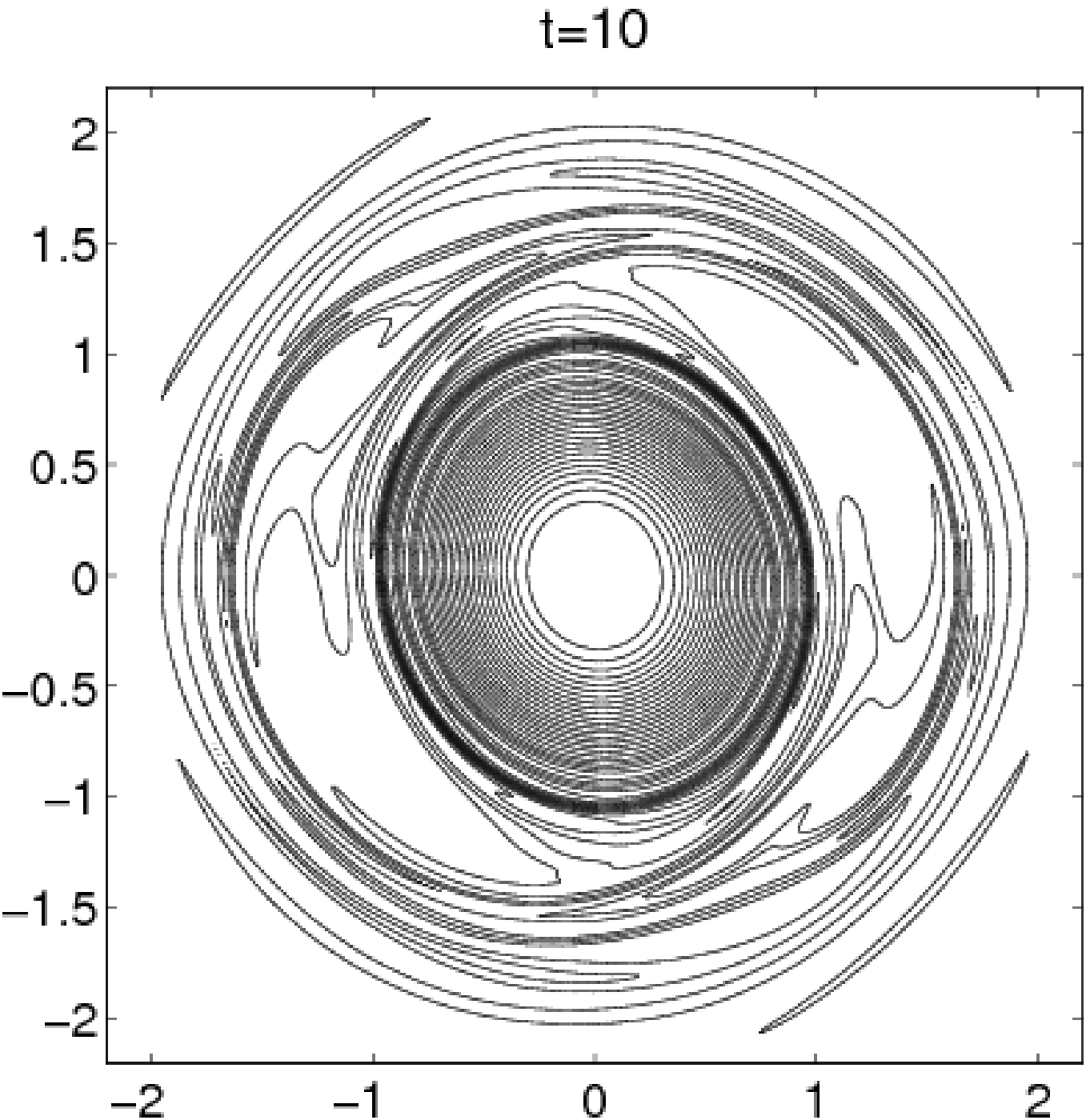}}
\end{minipage}
\caption{\label{fig:sumupn1} Contours of vorticity for Run4.}
\end{figure}

\begin{figure}
\centering
\begin{minipage}[c]{.7 \linewidth}
\centering
\scalebox{0.4}[0.4]{\includegraphics[width=\linewidth]{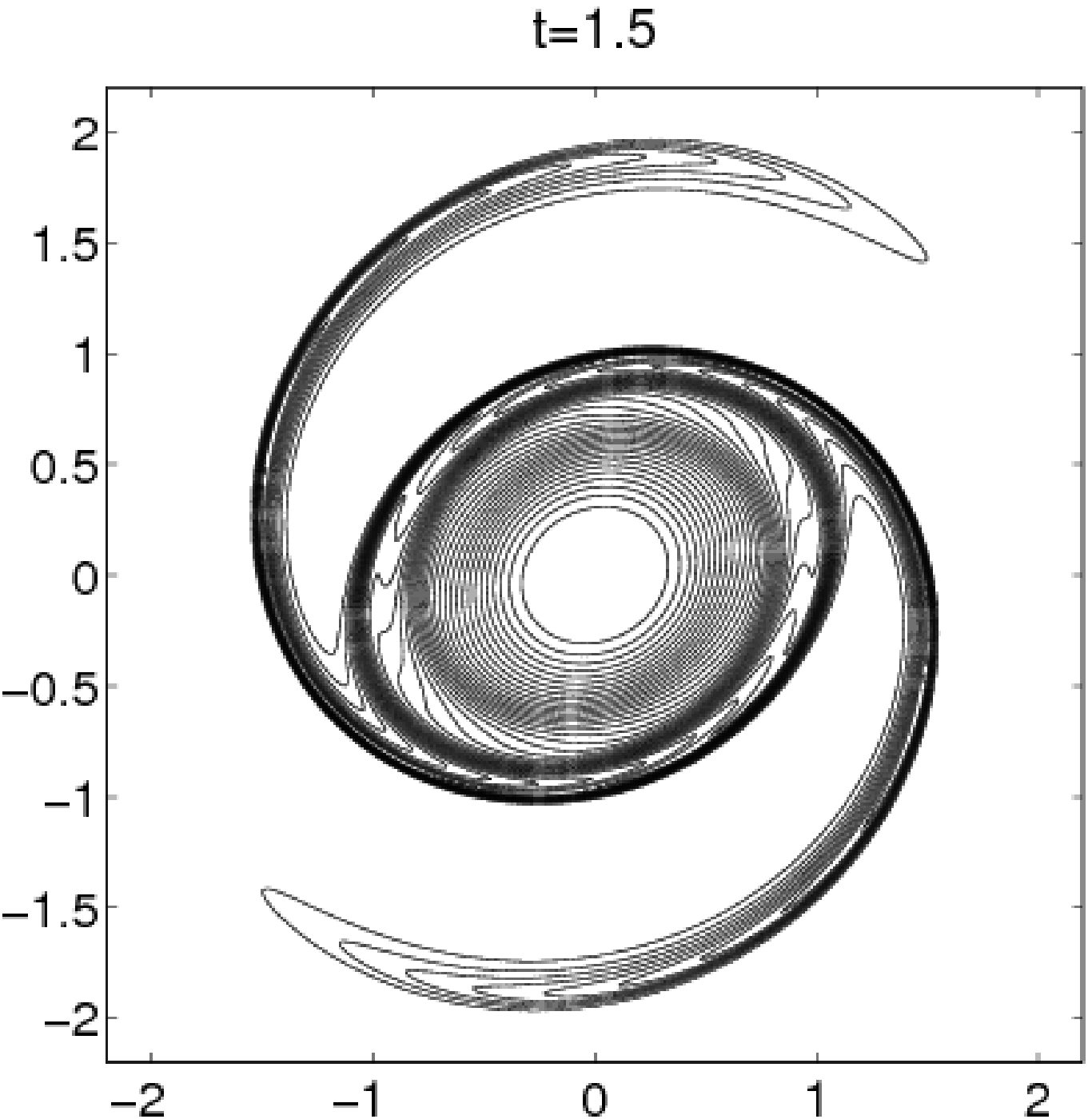}}
\scalebox{0.4}[0.4]{\includegraphics[width=\linewidth]{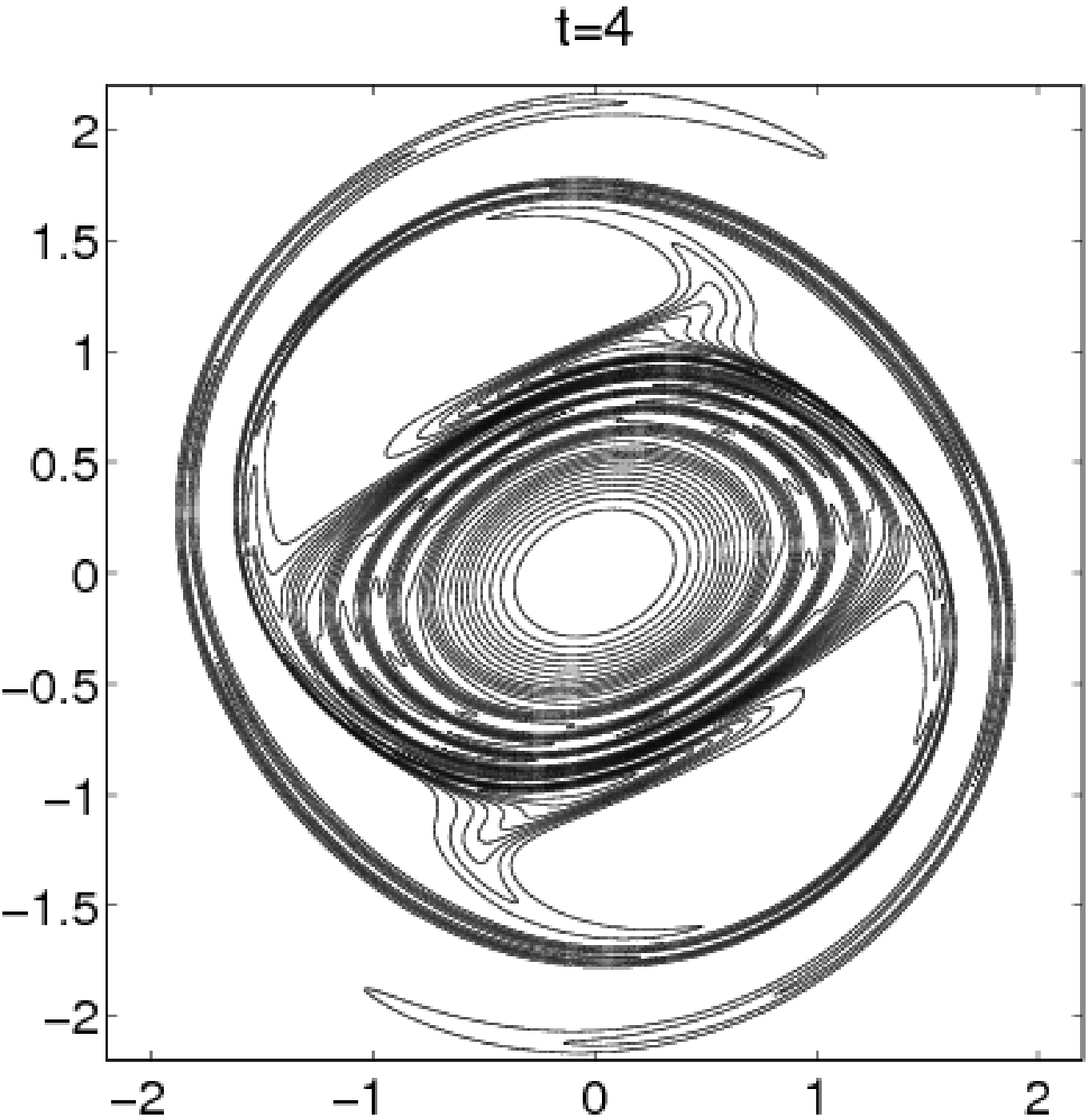}}
\scalebox{0.4}[0.4]{\includegraphics[width=\linewidth]{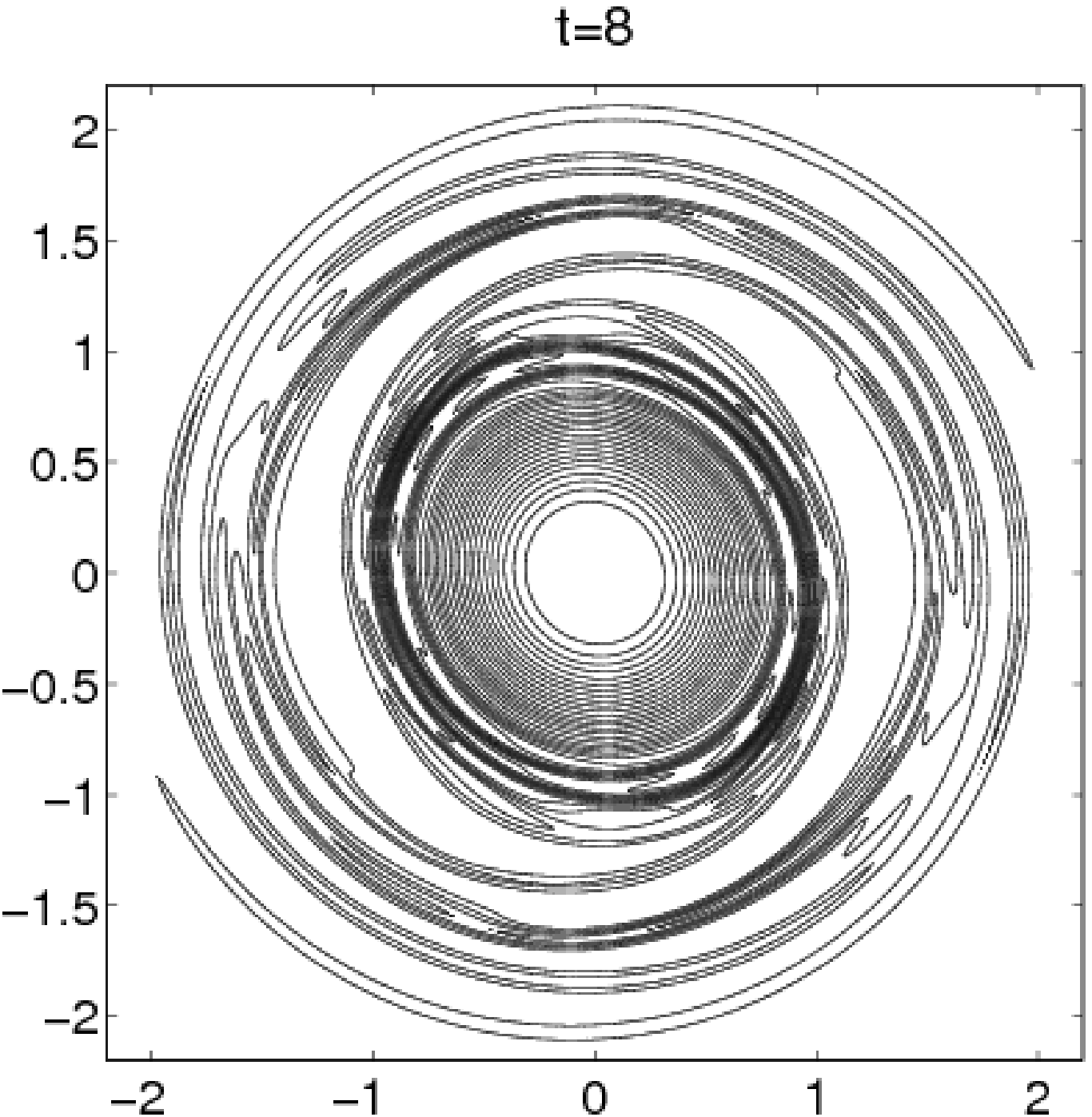}}
\scalebox{0.4}[0.4]{\includegraphics[width=\linewidth]{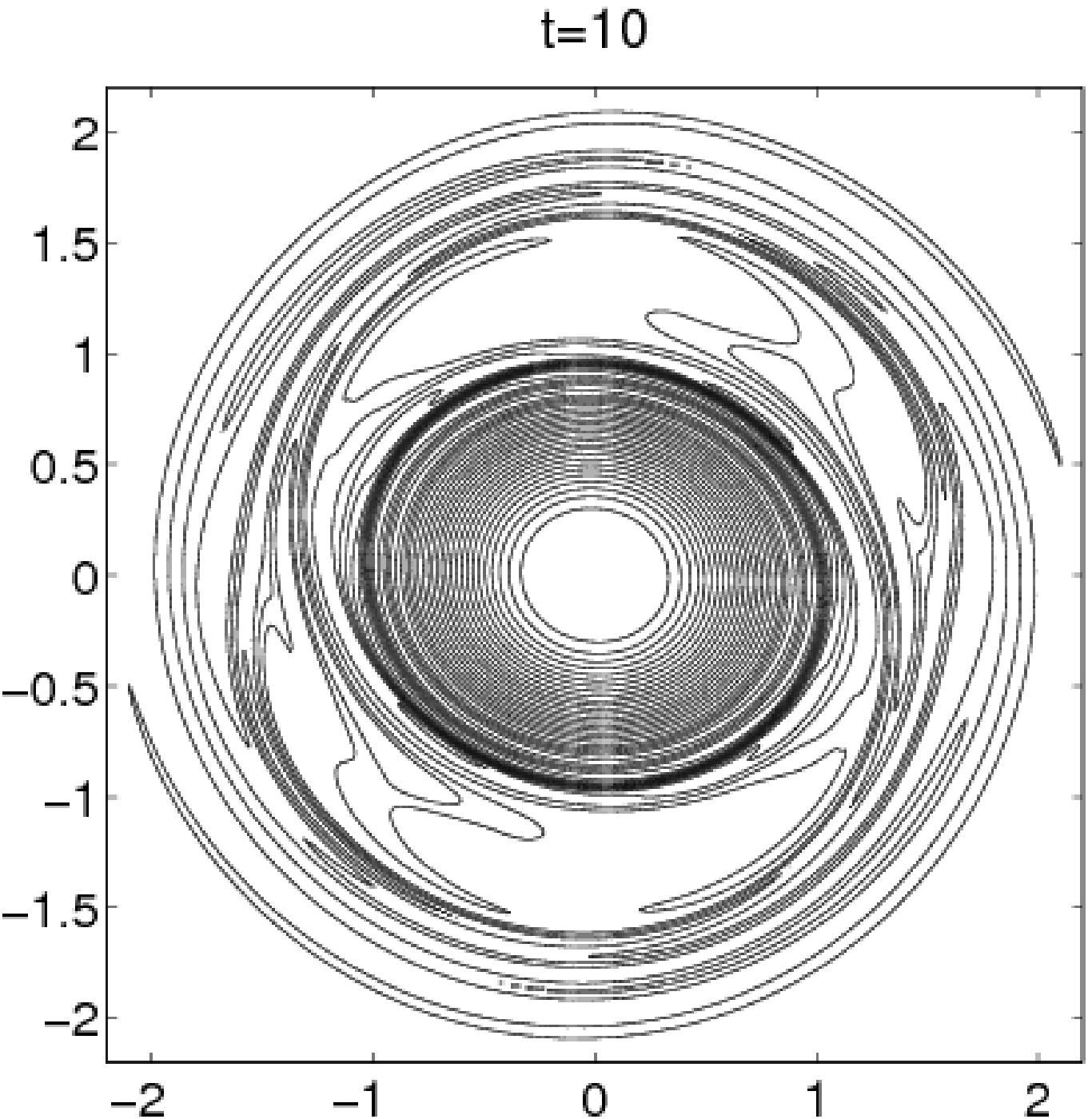}}
\end{minipage}
\caption{\label{fig:sumupn2} Contours of vorticity for Run5.}
\end{figure}

\subsection{Axisymmetrization of an isolated vortex}

In this subsection, we will conduct a study of the axisymmetrization of an isolated vortex based on that used in the 1987 paper by Melander, McWilliams, and Zabusky~\cite{McWilliams1987}. The governing equations are
\begin{equation}
\label{axisymm1}
\frac{\partial \omega }{\partial t} + u\frac{\partial \omega }{\partial x}
+ \upsilon\frac{\partial \omega }{\partial y} =  - \nu_h \Delta^2 \omega,
\end{equation}
\begin{equation}
\label{axisymm2}
\Delta \psi = -\omega, \; u=\frac{\partial \psi}{\partial y} , \; \upsilon=-\frac{\partial \psi}{\partial x}.
\end{equation}
Here, $\nu_h$ is hyperviscosity.

The initial conditions are an elliptical vortex with a smooth transition between rotational and irrotational fluid:
\begin{equation}
\label{mcwill}
\omega(x,y,0) = \left \{ \begin{array}{ll}
                         20 - 20 \exp \left[-\frac{\kappa}{r} \exp \left( \frac{1}{r-1} \right) \right]  & r < 1 \\
                         0 & r \geq 0 ,
                         \end{array}
                \right.
\end{equation}
where $r \equiv \sqrt{\frac{x^2}{2}+2y^2}$ and $\kappa=\frac{1}{2} e^2 ln(2)$. The similar question has been investigated by quite a few groups, and two numerical schemes are selected and compared here. One is the Fourier pseudospectral method which has high spectral accuracy but double periodical condition~\cite{McWilliams1987}, and another is the vortex method which is also implemented on the infinite domain~\cite{Platte2009,Koumoutsakos1997}.

The evolution of vortex is central symmetric, but not axisymmetric any more. Three simulations are performed:

\begin{description}
\item[Run1] uses $225 \times 225$ grids within $[-\pi, \pi] \times [-\pi, \pi]$, $\nu_h = 10^{-7}$, and the time step is $5.0 \times 10^{-4}$;

\item[Run2] uses $450 \times 450$ grids within $[-2\pi, 2\pi] \times [-2\pi, 2\pi]$, $\nu_h = 10^{-7}$, and the time step is $2.5 \times 10^{-4}$;

\item[Run3]  uses $400 \times 400$ grids within $[-\pi, \pi] \times [-\pi, \pi]$, $\nu_h = 0.3125 \times 10^{-7}$, and the time step is $1.5625 \times 10^{-4}$.
\end{description}

\vspace{0.1in}

Throughout the simulations, the region of our main interest, which covers all non-zero vorticity, is confined in the box of $[-2.2, 2.2] \times [-2.2, 2.2]$  (see Figs. \ref{fig:sumupn}). For Run1 and Run2, the shortest distances between grid points are almost identical ($\approx 0.02$). Fig. \ref{fig:ngrids} also shows that the gird points almost equally distribute near the center, so Run1 and Run2 can roughly be treated as simulations with the same resolution. Of course, the effective resolution of Run3 is the highest among the three simulations.

An easy self-checking technique is to study the time evolutions of maximum vorticity, which is conserved for an inviscid fluid. The relative errors of the maximum vorticity of the three runs are shown in Fig. \ref{fig:Merror}. It is obvious that Run3 has the highest precision, and the error is around $10^{-4}$ all the time. For the two compared runs, the maximum vorticity of Run2 is better conserved than Run1 throughout the whole simulations. Despite the similar effective resolutions, the spectral accuracy of our solver brings Run2 a higher precision than that of Run1. On the other hand, the corresponding error in the vortex methods is about ten times larger than our results (see, e.g. \cite{Koumoutsakos1997}).

It has been shown that for Fourier spectral methods, some artificial boundary conditions such as the sponge-layer are required to absorb the incoming vortex filaments~\cite{Mariotti1994}. Without any special treatment, the Fourier spectral method will cause about 17-degree overrotation at $t=1.65$ in the smaller domain simulation ($[-\pi, \pi] \times [-\pi, \pi]$), and about 3-degree overrotation for the larger domain $[-2\pi, 2\pi] \times [-2\pi, 2\pi]$~\cite{Platte2009}. However, no overrotation appears for Run1 and Run2 despite the different sizes of grid boxes because there is no boundary in our simulations. Also, a good agreement is reached when the vorticity evolutions of our simulations are compared with those of vortex method (Figs. \ref{fig:sdsize} \& \ref{fig:sumupn}). At the late stages of simulations, more and more filaments are generated, and it becomes a challenge for any numerical scheme to capture all these fine vortex structures. The related vorticity contours of Run3 are presented in Figs. \ref{fig:sumupn} and those very fine filaments are captured even after $t>8$. So, for the current problem, the Hermite spectral method has the merits of both numerical schemes having been compared: the high spectral accuracy as that of Fourier spectral methods, and the infinite domain computing as that of vortex methods.

Two compared simulations with the same $\nu_h$ and initial conditions as those of Run3 are performed by the Fourier Pseudo-spectral methods~\cite{Yin2004}:
\begin{description}
\item[Run4] uses $512 \times 512$ grids on the domain of $[-\pi, \pi] \times [-\pi, \pi]$;

\item[Run5] uses $1024 \times 1024$ grids on the domain of $[-2\pi, 2\pi] \times [-2\pi, 2\pi]$.

\end{description}
Except the phenomenon of overrotation, there is no big difference in the contour plots of Fourier and Hermite simulations until $t=4$. The overrotation in Run4 causes some vortex filaments to drift away from the view box at $t=8$, and some peripheral filaments break at $t=10$ (Figs. \ref{fig:sumupn1}). The large domain size in Run5 not only alleviates the overrotation, but also keeps the vortex filaments unbroken and well confined in the view box (Figs. \ref{fig:sumupn2}).

\begin{figure}
\centering
\scalebox{1}[1]{\includegraphics[width=\linewidth]{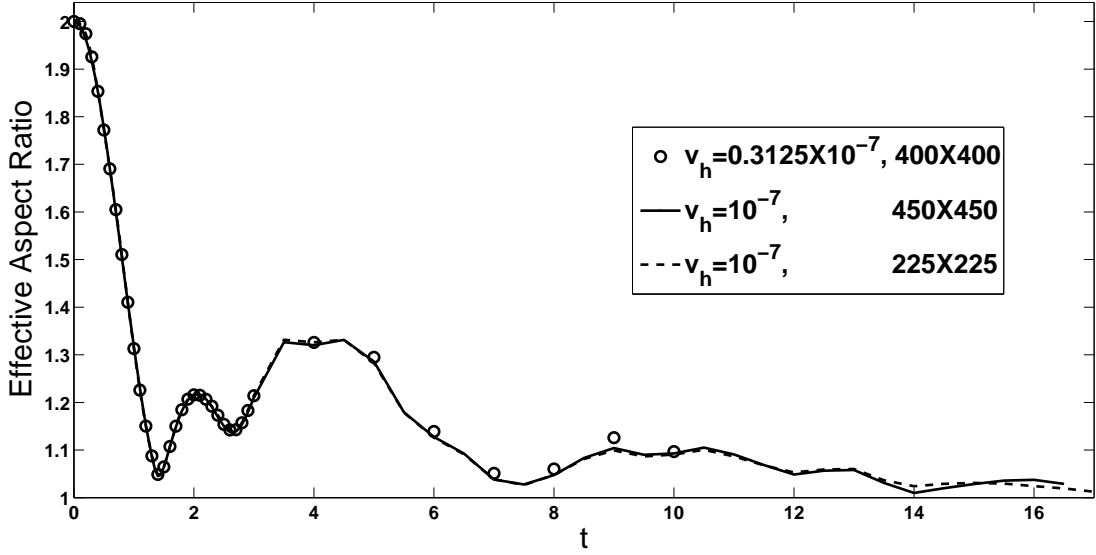}}
\caption{\label{fig:Lerror} Evolution of the effective aspect ration of the vorticity field from the three simulations in subsection 3.3.}
\end{figure}

The phenomenon of axisymmetrization can be quantified by considering the effective aspect ratio of vortical structures defined as
\begin{equation}
\label{Lamda_error}
\lambda=\sqrt{\frac{G+R}{G-R}},
\end{equation}
where $R^2 = D^2 + 4 G^2_{11}$, $D=G_{20}-G_{02}$, $G=G_{20}+G_{02}$, and $G_{mn}=\int\int \omega(x,y) x^m y^n dxdy$. Despite different $\nu_h$s and resolutions, the time evolutions of effective aspect ratios show good agreement for all three simulations in this subsection. Run3, which has a different $\nu_h$, begins to have some difference from the other two runs after $t=5$. There is almost no difference for the two compared runs until $t=13$. Again, the time evolutions of aspect ratio here are in good agreement with those of previous studies (e.g., \cite{Koumoutsakos1997}).

\vspace{0.2in}

Finally, two numerical schemes for the infinite domain can be compared.
\begin{itemize}
\item Vortex methods are developed as a grid-free methodology that is not limited by the fundamental smoothing effects associated with grid-based methods. The small scale and large scale are accurately simulated at the same time. They are especially well-suited to simulating filamentary motion, such as wisps of smoke, and in real-time simulations such as video games, because of the fine detail achieved by using minimal computation.

\item The Hermite spectral method is a grid-based method. If all non-zero vortices throughout simulations are confined in a small region, the spectral method has a higher accuracy than vortex methods. On the other hand, if some vortex filaments are drifted away from the view box, the Hermite spectral method requires a smaller scale factor to capture them. This means that the grid points are spread on a larger region, and more grid points are needed to guarantee the effective resolution within the region of our main interest. Moreover, a lot of computing time is wasted since vorticity is zero on most grid points.
\end{itemize}

\section{Conclusions}

To sum up, the Hermite spectral scheme presented in this paper can be used in solving 2D NS equations on the infinite domain. Of course, because of the definition of the Hermite function (Eq. (\ref{hermite})), the resultant solutions of this solver should decay exponentially as $(x,y) \rightarrow \infty$.  The primary variable in this paper is $\omega$, and the vorticity outside the box of grid points is almost zero, and thus the Hermite spectral methods apply. In the future, the same solver will be adopted to explore the Oseen vortex as a maximum entropy state of a two dimensional flow~\cite{Montgomery2011}. The Hermite spectral methods may also be used in exploring the phenomena of thermocapillary drop migration, where velocity far away from the drop is close to zero~\cite{Yin2008}.

\section*{Acknowledgements}
This project is supported by the NSF of China (Contract No. G11172308). I thank Prof. David C. Montgomery, Prof. Louis F. Rossi, Prof. William H. Matthaeus, and Dr. Hengbin An for their helpful suggestions.

\bibliographystyle{elsarticle-num}

\end{document}